\documentclass[nolineno,a4paper,USenglish,cleveref]{socg-lipics-v2021}
\pdfoutput=1
\usepackage{amsmath,amssymb}
\usepackage{cleveref}
\usepackage{csquotes}
\usepackage{xspace}
\usepackage{hyperref}
\usepackage{graphicx}

\newcommand{\TODO}[2][\empty]{{\normalfont\color{red}(\ifx #1\empty TODO:\else TODO #1:\fi\ {#2})}}

\newcommand{\ER}{\ensuremath{\exists\mathbb{R}}}

\newcommand{\class}[1]{\textsf{#1}} 
\newcommand{\lang}[1]{\textsf{#1}} 

\newcommand{\NP}{\class{NP}}
\newcommand{\SGE}{\lang {SGE}}
\newcommand{\ETR}{\lang {ETR}}
\newcommand{\SEFE}{\lang {SEFE}}
\newcommand{\PSPACE}{\class{PSPACE}}

\newcommand{\lt}{\ensuremath{\mathtt{l}}}
\newcommand{\rt}{\ensuremath{\mathtt{r}}}
\newcommand{\cl}{\ensuremath{\mathtt{c}}}

\graphicspath{{fig/}}

\graphicspath{{figs/}}

\title{On the Complexity of Simultaneous Geometric Embedding for Edge-Disjoint Graphs}
\author{Benedikt K\"unzel}{FernUniversität in Hagen, Germany}{benedikt.kuenzel@googlemail.com}{}{}
\author{Jonathan Rollin}{FernUniversität in Hagen, Germany}{jonathan.rollin@fernuni-hagen.de}{https://orcid.org/0000-0002-6769-7098}{}
\authorrunning{B. K\"unzel, J. Rollin}
\Copyright{B. K\"unzel, J. Rollin}
\hideLIPIcs

\begin{document}
\maketitle

\begin{abstract}
	Simultaneous Geometric Embedding (\SGE) asks whether, for a given collection of graphs on the same vertex set $V$, there is an embedding of $V$ in the plane that admits a crossing-free drawing with straightline edges for each of the given graphs.
	It is known that \SGE\ is \ER-complete, that is, the problem is  polynomially equivalent to deciding whether a system of polynomial equations and inequalities with integer coefficients has a real solution.
	We prove that \SGE\ remains \ER-complete for edge-disjoint input graphs, that is, for collections of graphs without so-called public edges.
	
	As an intermediate result, we prove that it is \ER-complete to decide whether a directional walk without repeating edges is realizable.
	Here, a directional walk consists of a sequence of not-necessarily distinct vertices (a walk) and a function prescribing for each inner position whether the walk shall turn left or shall turn right.
	A directional walk is realizable, if there is an embedding of its vertices in the plane such that the embedded walk turns according to the given directions.
	Previously it was known that realization is \ER-complete to decide for directional walks repeating each edge at most~336 times.
	
	This answers two questions posed by Schaefer [\textit{On the Complexity of Some Geometric Problems With Fixed Parameters}, JGAA 2021].
\end{abstract}

\section{Introduction}

The complexity status of many problems from computational geometry and graph drawing has been clarified in more detail in recent years.
While \NP-hardness has often been established earlier, membership with \NP\ seemed unlikely as solutions often require super-exponentially large coordinates~\cite{Sch09}.
In 2009, Schaefer~\cite{Sch09} introduced the complexity class \ER\ as the closure of
the problem of deciding the \emph{Existential Theory of the Reals} (\ETR) under polynomial-time reductions.
Here, \ETR\ is the set of true statements over the reals that can be written in the form $\exists(x_1,\ldots,x_n):\phi(x_1,\ldots,x_n)$, where $\phi$ is a Boolean formula over polynomial equations and inequalities with integer coefficients.
It is known that $\NP \subseteq \ER \subseteq \PSPACE$~\cite{Canny88,Sch09} and many geometric decision problems turned out to be \ER-complete (for examples see~\cite{AAM22, Cardinal15, Kyncl11}).

\begin{figure}
	\centering
	\includegraphics{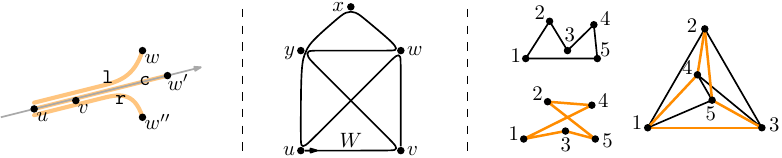}
	\caption{Left: Order types $\chi(u,v,w)=\lt$, $\chi(u,v,w')=\cl$, and $\chi(u,v,w'')=\rt$. Middle: A realization of the directional walk $u v^\lt y^\rt w^\lt x^\lt y^\lt u^\lt w^\rt v$. Right: Two edge-disjoint $5$-cycles on the same vertex set and a simultaneous geometric embedding.}
	\label{fig:intro-examples}
\end{figure}

Among these problems is also the question of \emph{Simultaneous Geometric Embedding} (\SGE), asking whether several graphs with a common vertex set admit a drawing with straightline edges on the same vertex locations in the plane such that each graph individually is drawn without crossings.
\Cref{fig:intro-examples} (right) shows an example of two simultaneously embedded graphs.
Cardinal and Kusters~\cite{CK15} describe a reduction from the \ER-complete problem \emph{Abstract Order Type Realizability} to Simultaneous Geometric Embedding.
Abstract Order Type Realizability, or the dual problem of stretchability of an arrangement of pseudolines, forms the base of many \ER-hardness proofs of geometric decision problems~\cite{Sch09} and we will describe this problem in more detail next.

Let $u$, $v$, and $w$ be distinct points in $\mathbb{R}^2$.
Then $w$ is either to the left of, or to the right of, or on the directed line from $u$ through $v$, see \cref{fig:intro-examples} (left).
We call this the \emph{order type} of the triple $(u,v,w)$ and denote it $\chi(u,v,w)\in\{\lt,\cl,\rt\}$,  where $\lt$ stands for left, $\cl$~for collinear, and~$\rt$ for right.
The \emph{order type} $\chi$ of a set of points $P\subset \mathbb{R}^2$ is then defined as $\chi: P^3 \rightarrow \{\lt,\cl,\rt\}$.
Given an abstract universe of points $U=\{u_1,\ldots,u_n\}$, an \emph{abstract order type} of $U$ is a mapping $\chi:U^3 \rightarrow \{\lt,\cl,\rt\}$.
We say that $\chi$ is \emph{realizable} if the points in $U$ can be mapped to points in $\mathbb{R}^2$ such that the order type of the resulting point set matches $\chi$.
The problem \emph{Abstract Order Type Realizability} asks whether a given abstract order type is realizable.
It follows from results of Mn\"ev~\cite{Mnev88} that this problem is \ER-complete.
Knuth defined a set of five axioms and called abstract order types satisfying these axioms \emph{CC-Systems} \cite{Knuth92}.
Knuth then showed that any order type of a set of points in $\mathbb{R}^2$ is a CC-System. It follows that all realizable abstract order types are CC-Systems.
Order types and CC-systems are special cases of chirotopes and oriented matroids, where realizability is also of fundamental interest~\cite{RZ04}.


Interesting questions arise when the order type is not prescribed for all triples of points, but only for a certain subset.
The corresponding realization problems are called \emph{Partial Order Type Realizability}.
A result of Schaefer~\cite[Theorem 1]{Sch21} implies that Partial Order Type Realizability remains \ER-complete, even for the set of instances where each vertex is contained in at most $\binom{131}{2}$ triples with prescribed order type\footnote{Schaefer's notion of \emph{thin}~\cite{Sch21} implies that each vertex has at most $131$ neighbors in the associated $3$-uniform hypergraph. Schaefer's result applies to even more restricted systems of prescribed order types and a more careful analysis might give a smaller bound on the degrees.}.
One of the motivations for studying partial order types are reductions of Partial Order Type Realizability to restricted variants of other \ER-complete geometric problems.
For instance, Schaefer used the result mentioned above to show \ER-completeness of deciding whether \emph{bounded-degree} graphs are segment intersection graphs -- a result that was known for graphs without degree constraints before~\cite{Matousek14}.
In the next paragraph, we describe another variant of partial order types, which is then used to show that a restricted variant of simultaneous geometric embedding is \ER-complete.
A discussion of further variants of partial order types is included in the conclusions in~\cref{sec:conclusions}.

Schaefer~\cite{Sch21} introduced the following \emph{Directional Walk Problem}.
A \emph{directional walk} $W$ consists of a sequence of not-necessarily distinct vertices $u_1,\ldots,u_t$ (a walk) and a function $d\colon [t-2]\to \{\lt, \rt\}$ prescribing directions.
The idea is that the walk turns in direction $d(i)$ after traveling the straightline segment from vertex $u_i$ to vertex $u_{i+1}$.
We write $V(W)$ to denote the set (not the sequence) of vertices in $W$ and $d_W$ to denote the function prescribing directions in $W$.
The pairs $\{u_i,u_{i+1}\}$ are called the edges of $W$ and $t$ is called the length of $W$ (note that $t$ might be much larger than $\lvert V(W)\rvert$ and much larger than the number of edges of $W$).
We also write $u_1$, $u_2^{d(1)}$, $u_3^{d(2)}$, \ldots, $u_t$ to describe a directional walk.

\Cref{fig:intro-examples} (middle) shows an example of a directional walk.
Note that a directional walk can be directly interpreted as a partial order type $\chi(u_i, u_{i+1}, u_{i+2}) = d_W(i)$, with $1\leq i \leq t-2$.

\begin{definition}[Directional Walk Problem~\cite{Sch21}]	
	 Given a directional walk $W$, decide whether $W$ is \emph{realizable}, that is, whether there is an embedding of $V(W)$ in the plane such that for each $i$, $1\leq i\leq t-2$, we have $\chi(u_i,u_{i+1},u_{i+2})= d_W(i)$.
\end{definition}

\begin{theorem}[Schaefer~\cite{Sch21}]\label{thm:walk-336times}
	The Directional Walk Problem is \ER-complete, even if each edge appears at most $336$ times in the walk.
\end{theorem}

Schaefer~\cite{Sch21} asked whether the problem remains  \ER-hard, even if the walk does not repeat any edges.
We answer this question in the affirmative.

\begin{theorem}\label{thm:walk-1time}
The Directional Walk Problem is \ER-complete for walks without repeated edges.
\end{theorem}

Note that we consider undirected edges and hence the walks in the above theorem are not allowed to revisit an edge in either direction.

We now formally introduce \emph{Simultaneous Geometric Embedding} (\SGE) which we already mentioned as an example of an \ER-complete problem above.
A \emph{geometric drawing} of a graph is a drawing where each edge is represented as a straightline segment.

\begin{definition}[\SGE
\protect\footnote{In our definition, the term \enquote{\SGE} refers to the problem with an arbitrary number of input graphs. In the literature the term is often used in case $\lvert \mathcal{G} \rvert = 2$ exclusively, while the general problem is called \enquote{$k$-\SGE\ with unbounded $k$}, where $k=\lvert \mathcal{G} \rvert$.}]
	Given a finite set $\mathcal{G}$ of graphs on the same vertex set $V$, decide whether there is an embedding of $V$ in the plane such that for each $G \in \mathcal{G}$ a geometric drawing of $G$ on that embedding is crossing-free.
	If such an embedding exists, then we say that the graphs in $\mathcal{G}$ can be \emph{simultaneously embedded}.
\end{definition}

\ER-hardness of \SGE\ follows from works of Kyncl~\cite{Kyncl11} and Gassner et al.~\cite{GJPSS06} (see~\cite{CK15}).
Using the result on partial order types mentioned above, Schaefer~\cite{Sch21} showed that \SGE\ is \ER-complete, even if each edge occurs in at most two input graphs.
In the language of \SGE, each edge that occurs in more than one input graph is called \emph{public}. 
If all public edges occur in all graphs, we say that the graphs in $\mathcal{G}$ share a \emph{Sunflower Intersection}.
The \emph{Sunflower Variant} of \SGE\ restricts the input to sets of graphs that share a Sunflower Intersection~\cite{Sch21}. 

We use our result from \cref{thm:walk-1time} above to prove that \SGE\ remains \ER-hard, even if the input graphs are edge-disjoint, answering a question of Schaefer.
In the language of \SGE\ this means that there are no public edges and so it is an extreme case of the sunflower variant.

\begin{theorem}\label{thm:sge}
\SGE\ is \ER-complete for edge-disjoint input graphs.
\end{theorem}

\paragraph*{Remark}

After we finished this paper, F\"orster et al.~\cite{FKMPTV23} published a preprint which independently proves \ER-completeness for \SGE\ on edge-disjoint input graphs.
They show that \SGE\ is \ER-complete for families of at most 58 edge-disjoint input graphs.

\paragraph*{Related Work on \SGE}

There are many different variants of drawing graphs simultaneously.
A systematic study of simultaneous embeddings was initiated by Brass et al.~\cite{BrassEtAl07} and surveys are given by Bl\"asius et al.~\cite{BKR13} and Rutter~\cite{Rutter20}.

Concerning the setting studied in this paper, Brass et al.~\cite{BrassEtAl07} showed that two paths always admit a simultaneous geometric embedding while there are three paths that do not admit such an embedding.
Angelini et al.~\cite{AGKN12} described a path and an edge-disjoint tree that do not admit a simultaneous geometric embedding.
\SGE\ is \ER-complete for collections of $9360$ paths or collections of $240$ graphs \cite{Sch21} and the complexity status is open in case of just two input graphs.

In our setting we have a fixed common labeling of the vertices across all input graphs.
If for each graph the embedded vertex set may be permuted we obtain a version called \emph{\SGE\ without mapping}.
Recently, Steiner~\cite{Steiner23} proved the existence of a collection of $O(\log n)$ many $n$-vertex graphs that do not admit a simultaneous geometric embedding without fixed mapping.
It is open whether any two graphs do admit such a simultaneous embedding and the complexity appears to have not been investigated.

Yet another variant is called \emph{Simultaneous Embedding with Fixed Edges} (\SEFE).
In this setting the vertex labeling is fixed across all input graphs but the edges are not required to be straightline segments anymore.
Instead, each edge must be drawn on the same curve in the plane for each graph it occurs in.
The surveys mentioned above present many results for this variant.
For example, Frati~\cite{Frati06} showed that any tree and any planar graph do admit a \SEFE\ while there are two outerplanar graphs that do not.
There are polynomial-time algorithms checking whether two graphs with a $2$-connected intersection graph admit a \SEFE~\cite{ABFPR12,HJL13} but it is unknown whether such an algorithm exists for arbitrary pairs of graphs.
Angelini et al.~\cite{ALN15} showed that the problem becomes \NP-complete for three or more input graphs, even if the graphs share a sunflower intersection that is a star.
Further polynomially tractable as well as hard instances were identified by Fink et al.~\cite{FPR23}.
Note that edge-disjoint graphs always admit a \SEFE\ as every planar graph can be drawn in a planar way with arbitrary vertex locations~\cite{PW01} which is in stark contrast to our \cref{thm:sge} above.
This also shows that arbitrary graphs can be drawn simultaneously, if each graph's edges can be drawn on individual curves.

Finally, we mention geometric thickness.
A graph has geometric thickness at most $k$ if it can be partitioned into $k$ subgraphs that admit a simultaneous geometric embedding.
It is known that deciding whether a graph has geometric thickness $2$ is \NP-hard~\cite{DGM16} and contained in \ER\ but membership with \NP\ or \ER-hardness remain open~\cite{Sch21}.
\footnote{After we finished this paper, F\"orster et al.~\cite{FKMPTV23} published a preprint claiming \ER-hardness of deciding whether a multigraph has geometric thickness at most 57.}

\paragraph*{Notation}

Let $\lnot\lt = \rt$, $\lnot\rt = \lt$, and $\lnot\cl =\cl$.
We will use the identity $\chi(u,v,w)=\lnot \chi(w,u,v)$ several times which holds for any three points $u, v, w \in \mathbb{R}^2$.
This might be immediate from the geometry of the plane or, for instance, is the first axiom of CC-systems~\cite{Knuth92}.

\section{Proof of \cref{thm:walk-1time}}

\begin{figure}
	\centering
	\includegraphics{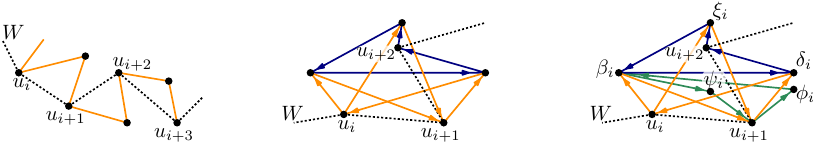}
	\caption{Left: A walk (dotted black) with some hooks (solid orange). Middle: For each edge $\{u_i,u_{i+1}\}$ three hooks are placed (orange) and a walk (blue) ensuring that $u_{i+2}$ is placed in the triangle formed by the dummy vertices of the hooks. Right: The hooks and the placement of $u_{i+2}$ are combined to a triangle gadget via two additional dummy vertices.}
	\label{fig:walk-replacement}
\end{figure}

We will use \cref{thm:walk-336times} which states that the Directional Walk Problem is \ER-complete.
This directly implies that the restricted variant lies in \ER.
We start with a brief outline of the \ER-hardness proof.
Given a directional walk $W$ with vertex sequence $u_1,\ldots,u_t$ we shall construct a new directional walk $W'$ using the vertices of $W$ and additional dummy vertices such that $W'$ does not repeat any of its edges and is realizable if and only if $W$ is realizable.
To achieve the latter property, $W'$ enforces the same partial order type on $V(W)$ as prescribed by $W$.
For each triple $u_i$, $u_{i+1}$, $u_{i+2}$ of vertices along $W$ we create several dummy vertices $x$ such that in any realization of $W$ we have $\chi(u_i,u_{i+1},x)=d_W(i)$.
As we do not want to use the edge $\{u_i, u_{i+1}\}$ in $W'$ we use the equivalent order type $\chi(u_i,x,u_{i+1})=\lnot d_W(i)$ and walk along $u_i$, $x$, $u_{i+1}$ in $W'$ instead, or we use the order type $\chi(u_{i+1},x,u_i)= d_W(i)$ and walk along $u_{i+1}$, $x$, $u_{i}$ in $W'$, see \cref{fig:walk-replacement} (left).
We call such a part of $W'$ a \emph{hook}.
After placing three such hooks at $\{u_i,u_{i+1}\}$ we proceed with $W'$ in such a way that $u_{i+2}$ is placed in the triangle formed by the dummy vertices in the hooks.
This will enforce the correct order type of $(u_i, u_{i+1}, u_{i+2})$, see \cref{fig:walk-replacement} (middle).

We give a precise definition of $W'$ next.
For each $i$, $1\leq i\leq t-2$, we use five dummy vertices $\beta_i$, $\delta_i$, $\xi_i$, $\phi_i$, and $\psi_i$.
The first three are used in hooks and the latter two are used to combine the hooks and the placement of $u_{i+1}$ to a single walk, starting at $u_i$ and ending at $u_{i+1}$.
The \emph{triangle gadget} for $i$ is the following directional walk, where $d=d_W(i)$:
\begin{align}&\hphantom{u_i, \beta_i^{\lnot d},\ }\overbrace{\hphantom{u_{i+1}^{d},  \delta_i^{d}, u_i^{\lnot d}}}^\text{hook (ii)}\nonumber\\[-8pt]
W_i = &\underbrace{u_i, \beta_i^{\lnot d}, u_{i+1}^{d}}_{\text{hook (i)}}, \delta_i^{d}, \underbrace{u_i^{\lnot d}, \xi_i^{\lnot d}, u_{i+1}^d}_\text{hook (iii)}, \phi_i^d, \underbrace{\beta_i^{\lnot d}, \delta_i^{d}, u_{i+2}^{\lnot d}, \xi_i^{d}, \beta_i^d}_\text{$u_{i+2}$ in triangle (iv)}, \psi_i^{\lnot d}, u_{i+1}\label{eq:tri-gadget}
\end{align}

\Cref{fig:walk-replacement} (right) shows an example of a triangle gadget.
Now, $W'$ is defined as the concatenation of $W_1,\ldots,W_{t-2}$ where for each $i$, with $1< i \leq t-2$, the first vertex of $W_i$ is identified with the last vertex of $W_{i-1}$ and the first occurrence of $u_i$ in $W_i$ receives direction $\lnot d_W(i)$.
For example, assuming $d_W(1)=\lt$ and $d_W(2)=\rt$, the beginning of $W'$ is
\[W' = \underbrace{u_1, \beta_1^{\rt}, u_{2}^{\lt}, \delta_1^{\lt}, u_1^{\rt}, \xi_1^{\rt}, u_{2}^\lt, \phi_1^\lt, \beta_1^{\rt}, \delta_1^{\lt}, u_{3}^\rt, \xi_1^{\lt}, \beta_1^\lt, \psi_1^\rt, u_{2}^\lt}_{W_1}, \beta_2^{\lt}, u_{3}^{\rt}, \delta_2^{\rt}, u_2^{\lt}, \xi_2^{\lt}, u_{3}^\rt, \ldots\]

By construction, there are no repeated edges in $W'$.
Moreover, the walk $W'$ can be constructed in polynomial time (with respect to the length of $W$) as each occurrence of an edge in $W$ is replaced by a walk of constant length.
It remains to show that $W'$ is realizable if and only if $W$ is realizable.

If $W'$ is realizable, consider some realization of $W'$.
We will show that for each $i$, $1\leq i\leq t-2$, we have $\chi(u_i,u_{i+1},u_{i+2})=d_W(i)$ in this realization.
Due to the hooks in $W_i$ we have
\begin{align*}
	\chi(u_i,u_{i+1},\beta_i) &= \lnot\chi(u_i,\beta_i,u_{i+1}) \overset{\text{(i) in \labelcref{eq:tri-gadget}}}{=} d_W(i),\\
	\chi(u_i,u_{i+1},\delta_i) &= \chi(u_{i+1},\delta_i,u_i)\overset{\text{(ii) in \labelcref{eq:tri-gadget}}}{=} d_W(i),\\
	\chi(u_i,u_{i+1},\xi_i) &= \lnot\chi(u_i,\xi_i,u_{i+1}) \overset{\text{(iii) in \labelcref{eq:tri-gadget}}}{=} d_W(i).
\end{align*}
That is, the dummy vertices $\beta_i$, $\delta_i$, and $\xi_i$ are placed on the $d_W(i)$-side of the straight line from $u_i$ through $u_{i+1}$.
The latter part of the triangle gadget gives
\begin{align*}
	\chi(\beta_i,\delta_i,u_{i+2}) &\overset{\text{(iv) in \labelcref{eq:tri-gadget}}}{=} d_W(i),\\
	\chi(\delta_i,\xi_i,u_{i+2}) &= \lnot\chi(\delta_i,u_{i+2},\xi_i) \overset{\text{(iv) in \labelcref{eq:tri-gadget}}}{=}  d_W(i),\\
	\chi(\xi_i,\beta_i,u_{i+2}) &= \chi(u_{i+2},\xi_i,\beta_i) \overset{\text{(iv) in \labelcref{eq:tri-gadget}}}{=}  d_W(i).
\end{align*}
This implies that $u_{i+2}$ is placed in the interior of the triangle formed by $\beta_i$, $\delta_i$, and $\xi_i$.
Hence, $\chi(u_i,u_{i+1},u_{i+2})=d_W(i)$ as desired.

Now assume that $W$ is realizable and consider some realization of $W$.
We will construct a realization of $W'$ by positioning the dummy vertices, processing $W_1, \ldots, W_{t-2}$ one after the other in that order.
For each $i$, $1\leq i\leq t-2$, we have $\chi(u_i,u_{i+1},u_{i+2})=d_W(i)$ in the realization of $W$.
We now place the dummy vertices $\beta_i$, $\delta_i$, $\xi_i$, $\phi_i$, and $\psi_i$.
First, we apply an affine transformation of the plane to all points constructed so far such that
\begin{itemize}
	\item if $d_W(i) = \lt$, then $u_i$ is mapped to the point $(4,-2)$, $u_{i+1}$ is mapped to the point $(4,2)$, and $u_{i+2}$ is mapped to the point $(0,0)$,
	
	\item if $d_W(i) = \rt$, then $u_i$ is mapped to the point $(-4,-2)$, $u_{i+1}$ is mapped to the point $(-4,2)$, and $u_{i+2}$ is mapped to the point $(0,0)$.
\end{itemize}

Observe that the chosen transformation preserves the order types of all triples of points constructed so far, as it preserves the order type of $\{u_i, u_{i+1}, u_{i+2}\}$.
Now we place the dummy vertices like in \cref{fig:walk-dummy-placement}, depending on the values of $d_W(i)$ and $d_W(i+1)$:

\begin{figure}
	\centering
	\includegraphics{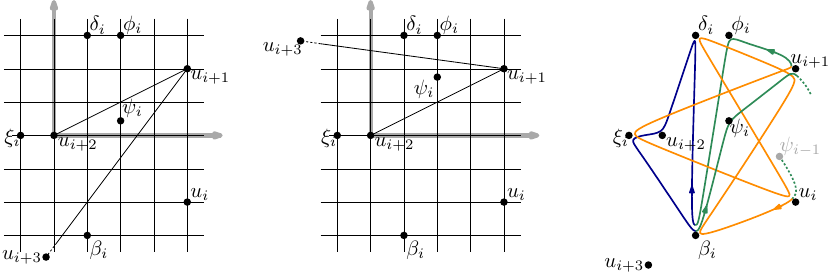}
	\caption{Left/Middle: Placement of the dummy vertices in the cases $d_W(i)=\lt$ and $d_W(i+1)=\lt$ (left) and $d_W(i+1)=\rt$ (middle). The thin lines represent the integer grid with origin at $u_{i+2}$. The positions of $u_i$, $u_{i+1}$, and $u_{i+2}$ are fixed by an affine transformation. The position of	$u_{i+3}$ in the figure is just sketched for illustrating the placement of $\psi_i$: $u_{i+3}$ is shown on the correct side of the line through $u_{i+1}$ and $u_{i+2}$ but its actual position on that side might differ.
	The placement in case $d_W(i)=\rt$ is obtained by mirroring the points along the y-axis.
	Right: The walk $W'$ in case $d_W(i)=\lt$ and $d_W(i+1)=\lt$.
	}
	\label{fig:walk-dummy-placement}
\end{figure}

\begin{itemize}
	\item If $d_W(i)=\lt$, then $\beta_i$ is placed at $(1,-3)$, $\delta_i$ is placed at $(1,3)$, $\xi_i$ is placed at $(-1,0)$, and $\phi_i$ is placed at $(2,3)$. The point $\psi_i$ is chosen with x-coordinate $2$ and non-negative y-coordinate (possibly not an integer) such that it is in the interior of the triangle formed by $u_{i+1}$, $u_{i+2}$, and $u_{i+3}$.
	
	\item If $d_W(i)=\rt$, then $\beta_i$ is placed at $(-1,-3)$, $\delta_i$ is placed at $(-1,3)$, $\xi_i$ is placed at $(1,0)$, and $\phi_i$ is placed at $(-2,3)$. The point $\psi_i$ is chosen with x-coordinate $-2$ and non-negative y-coordinate (possibly not an integer) such that it is in the interior of the triangle formed by $u_{i+1}$, $u_{i+2}$, and $u_{i+3}$.
\end{itemize}

Finally, we apply a small perturbation to ensure that the point set is in general position while preserving the order types of non-collinear triples of points.
It can be readily checked that this placement indeed realizes $W'$ by following the walk in \cref{fig:walk-dummy-placement} (right).
The placement of $\psi_i$ in the triangle formed by $u_{i+1}$, $u_{i+2}$, and $u_{i+3}$ (and the placement of $\beta_{i+1}$ below $u_{i+1}$) ensures that $\chi(\psi_i, u_{i+1}, \beta_{i+1})=\lnot d_W(i+1)$.

Altogether, we have a polynomial-time reduction that, for each directional walk $W$, gives a directional walk $W'$ such that $W'$ does not repeat any edges and $W'$ is realizable if and only if $W$ is realizable.
This finishes the proof, since deciding whether a directional walk is realizable is \ER-complete by \cref{thm:walk-336times}.

\section{Proof of \cref{thm:sge}}

We shall give a polynomial-time reduction from the Directional Walk Problem to \SGE.
To prove \ER-hardness of the restricted variant of \SGE, we restrict the input of the reduction to directional walks without repeating edges and receive \SGE-instances with edge-disjoint input graphs.
The reduction includes the vertices of the input directional walk $W$ in the vertex set of the \SGE-instance such that, if the instance admits a simultaneous embedding, then the embedding of $V(W)$ in any such embedding virtually realizes $W$.
The \SGE-instance contains, for each position $i$ in the walk prescribing some order type, an individual graph $G_i$ that is responsible for realizing that order type.
Additionally, there is a frame graph $G_\text{frame}$ ensuring that the graphs $G_i$ get embedded in a specific way.

Given a directional walk $W$ with vertex sequence $u_1,\ldots,u_t$ the reduction is precisely defined as follows.
The vertex set of the \SGE-instance is the disjoint union $V=\cup_{i=1}^{t-2} V_i \cup V_\text{frame} \cup V(W) \cup \{u'_j\colon u_j\in V(W)\}$ where
\[V_i =  \{a_i, a'_i, b_i, b'_i, c_i, c'_i, d_i, d'_i, e_i, e'_i, f_i, f'_i\} \text{ and } V_\text{frame} =   \{p, p', x, y, z\}.\]

\begin{figure}
	\centering
	\includegraphics{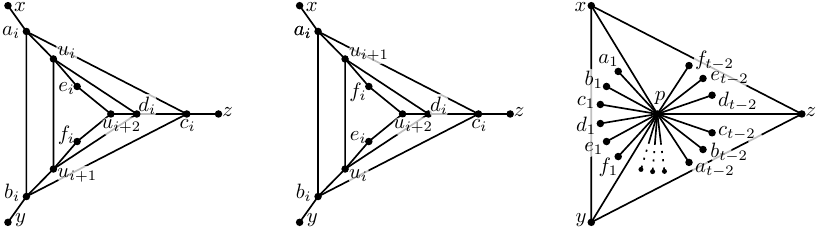}
	\caption{Left: The graph $\widetilde{G}_i$ in case $d_W(i)=\lt$. Middle: The graph $\widetilde{G}_i$ in case $d_W(i)=\rt$ (only the vertices $u_i$, $u_{i+1}$, $e_i$, and $f_i$ switched locations). Right: The graph $G_\text{frame}$ (primed vertices omitted).}
	\label{fig:SGE-gadget}
\end{figure}

We define the graphs $G_i$ in two steps.
For each $i$, with $1\leq i\leq t-2$, let $\widetilde{G}_i$ denote graph given in \cref{fig:SGE-gadget} (left or middle):
its vertex set is $\widetilde{V}_i = \{a_i, b_i, c_i, d_i, e_i, f_i, x, y, z\}$ and its edge set is the disjoint union $E(\widetilde{G}_i) = \widetilde{E}^i_\text{frame} \cup \widetilde{E}_i$ where
\begin{align*}
	\widetilde{E}^i_\text{frame} & = \{xa_i, yb_i, zc_i, a_ib_i, a_ic_i, b_ic_i\}\\
	\intertext{and in case $d_W(i)=\lt$}
	\widetilde{E}_i & = \{a_iu_i, b_iu_{i+1}, c_id_i, u_iu_{i+1}, u_id_i, u_{i+1}d_i, u_ie_i, u_{i+1}f_i, u_{i+2}d_i, u_{i+2}e_i, u_{i+2}f_i\},\\
	\intertext{and in case $d_W(i)=\rt$}
	\widetilde{E}_i & = \{a_iu_{i+1}, b_iu_{i}, c_id_i, u_iu_{i+1}, u_id_i, u_{i+1}d_i, u_ie_i, u_{i+1}f_i, u_{i+2}d_i, u_{i+2}e_i, u_{i+2}f_i\}.
\end{align*}

That is, the graph $\widetilde{G}_i$ consists of edges $xa_i$, $yb_i$, and $zc_i$, a triangle on $\{a_i, b_i, c_i\}$, a copy of $K_4$ on $\{u_i, u_{i+1}, u_{i+2}, d_i\}$ where the edge $u_iu_{i+2}$ is subdivided by $e_i$ and the edge $u_{i+1}u_{i+2}$ is subdivided by $f_i$ and, in case  $d_W(i)=\lt$, further edges $a_iu_i$, $b_iu_{i+1}$, $c_id_i$, and, in case  $d_W(i)=\rt$, further edges $a_iu_{i+1}$, $b_iu_{i}$, $c_id_i$.
Let $\widetilde{G}'_i$ denote a copy of $\widetilde{G}_i$ where each vertex label is primed, that is, $V(\widetilde{G}'_i) =\{a'_i, b'_i, c'_i, d'_i, e'_i, f'_i, x', y', z'\}$.
The graph $G_i$ is now obtained from the disjoint union of $\widetilde{G}_i$ and $\widetilde{G}'_i$ by identifying $x$ and $x'$, $y$ and $y'$, as well as $z$ and $z'$ to single vertices $x$, $y$, and $z$.
Additionally, all vertices from $V$ that are not yet present in $G_i$ are added as isolated vertices.
So the vertices $p$ and $p'$, all vertices in $V_j$, with $j\neq i$, and all vertices $u_j$, $u'_j$, with $j\not\in\{i,i+1,i+2\}$, are isolated in $G_i$.
Hence, the only edges shared by $G_i$ and $G_j$, with $i\neq j$, might be $u_iu_{i+1}$ and $u'_iu'_{i+1}$.
This, however, happens only if $\{u_i, u_{i+1}\}=\{u_j, u_{j+1}\}$ (as sets) which is impossible if the walk $W$ does not repeat any edges.

Further, let $G_\text{frame}$ denote the graph illustrated in \cref{fig:SGE-gadget} (right, primed vertices omitted) with vertex set $V$ consisting of a triangle on $\{x, y, z\}$, all edges between $p$ and $\bigcup_{i=1}^{t-2}\widetilde{V}_i \cup \{x,y,z\}$, and all edges between $p'$ and $\bigcup_{i=1}^{t-2}\{a'_i, b'_i, c'_i, d'_i, e'_i, f'_i\} \cup \{x,y,z\}$.
The graph $G_\text{frame}$ does not share any edges with any of the graphs $G_1,\ldots,G_{t-2}$ by construction.
As $\lvert V\rvert<14t$, the reduction can be carried out in polynomial time in $t$.
We claim that $G_1, \ldots, G_{t-2}$, and $G_\text{frame}$ can be simultaneously embedded if and only if the directional walk $W$ is realizable.

First assume that $G_1, \ldots, G_{t-2}$, and $G_\text{frame}$ can be simultaneously embedded.
That is, there is an embedding of $V$ such that for each of the graphs a geometric drawing on that embedding is crossing-free.
We will prove that we can derive an embedding of $V(W)\subseteq V$ that is a realization of $W$.
Consider the triangle $\Delta$ on $x$, $y$, and $z$ in $G_\text{frame}$ and the vertices $p$ and $p'$.
Due to the edges between $\{p,p'\}$ and $\{x,y,z\}$ in the crossing-free embedding of $G_\text{frame}$, one of $p$ and $p'$ is embedded in the interior of $\Delta$ while the other is embedded in the exterior.
If $p$ is in the exterior, then we swap all primed vertices with their corresponding non-primed vertices (while fixing $x$, $y$, $z$).
The swapped embedding still admits a simultaneous embedding of the graphs as this mapping is an isomorphism for each of $G_1,\ldots,G_{t-2}$ and $G_\text{frame}$.
So we assume that $p$ is embedded in the interior of $\Delta$.
Let $\widetilde{V}=\bigcup_{i=1}^{t-2} \widetilde{V}_i$ denote the set of all non-primed vertices, except $x$, $y$, $z$.
Then all vertices from $\widetilde{V}$ are embedded in the interior of $\Delta$ as well, as they are adjacent to $p$ in $G_\text{frame}$.
We reflect the plane, if necessary, to ensure that $\chi(x,y,z)=\lt$.
Note that the reflected embedding of $V$ still admits a simultaneous embedding of the graphs and still has all vertices from $\widetilde{V}$ in the interior of $\Delta$.

\begin{figure}
	\centering
	\includegraphics{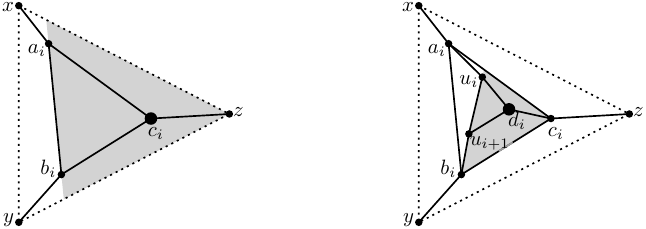}
	\caption{The gray areas shows all possible locations of $c_i$ (left), respectively $d_i$ (right).}
	\label{fig:SGE-ordertypes}
\end{figure}

We now show that for each $i$, with $1\leq i\leq t-2$, we have $\chi(u_i, u_{i+1}, u_{i+2})=d_W(i)$ in the embedding, that is, the embedding of $V(W)\subseteq \widetilde{V}$ realizes $W$.
To prove this, consider the path $x$, $a_i$, $b_i$, $y$ in $G_i$ that lies within $\Delta$.
Since $c_i$ is in the interior of $\Delta$ and adjacent to $a_i$, $b_i$, and $z$ we have $\chi(a_i,b_i,c_i)=\lt$ as $\chi(x,y,z)=\lt$ and as $G_i$ is drawn without crossings, see \cref{fig:SGE-ordertypes} (left).
Since all vertices from $\widetilde{V}$ are in the interior of $\Delta$, also the triangle $T$ on $\{d_i, u_i, u_{i+1}\}$ in $G_i$ is embedded within $\Delta$.
As it has an edge to each of $a_i$, $b_i$, and $c_i$ in the crossing-free drawing of $G_i$, $T$ is embedded in the interior of the triangle on $\{a_i, b_i, c_i\}$.
With a similar argument as above we see that $\chi(u_i,u_{i+1},d_i) = d_W(i)$:
If $d_W(i)=\lt$, we have a path $a_i$, $u_i$, $u_{i+1}$, $b_i$ in $G_i$ that lies within the triangle on $\{a_i, b_i, c_i\}$ and, if $d_W(i)=\rt$, we have a path $a_i$, $u_{i+1}$, $u_{i}$, $b_i$ in $G_i$, see \cref{fig:SGE-ordertypes} (right).
The case distinction comes from the differences in the construction of $G_i$ (respectively $\widetilde{G}_i$) depending on the value of $d_W(i)$.
In both cases, $d_i$ is adjacent to $u_i$, $u_{i+1}$, and $c_i$ which implies that $\chi(u_i,u_{i+1},d_i) = d_W(i)$.
Finally observe that $u_{i+2}$ lies in the interior of the triangle $T$ due to its connections to $d_i$, $u_i$ (via $e_i$), and $u_{i+1}$ (via $f_i$).
Hence $\chi(u_i,u_{i+1}, u_{i+2}) = \chi(u_i,u_{i+1},d_i) = d_W(i)$.
Altogether, this shows that the embedding of $V(W)$ realizes $W$.

Next assume that the directional walk $W$ is realizable.
We need to show that the graphs $G_1, \ldots, G_{t-2}$, and $G_\text{frame}$ can be simultaneously embedded.
To this end, we will specify the locations of the remaining vertices in $V$.
Consider an embedding of $V(W)$ that realizes $W$.
We apply an affine transformation of the plane, if necessary, to ensure that the embedded vertices lie within the unit square centered at the origin of the plane.
The transformed embedding still realizes $W$.
First place the vertex $x$ at position $(-1,2)$, the vertex $y$ at position $(-1,-2)$, and $z$ at position $(1,0)$.
Then $\chi(x,y,z)=\lt$ and the embedded vertices from $V(W)$ are in the interior of the triangle $\Delta$ formed by $x$, $y$, and $z$.
We proceed with the remaining non-primed vertices which will be placed in the interior of the triangle $\Delta$.
For each $i$, with $1\leq i\leq t-2$, we place the vertices from $\widetilde{V}_i$ (the non-primed vertices of $G_i$) in the interior of $\Delta$ as follows.
The vertices $e_i$ and $f_i$ are placed in the interior of the segments $u_{i+2}u_i$ and $u_{i+2}u_{i+1}$, respectively.
The vertex $d_i$ is placed close to to $u_{i+2}$ such that $u_{i+2}$ is in the interior of the triangle $T$ formed by $u_i$, $u_{i+1}$, and $d_i$.
So far, all edges of $G_i$ induced by already embedded vertices are crossing-free.
Then we place the vertices $a_i$, $b_i$, and $c_i$ such that
\begin{itemize}	
	\item they are in the interior of $\Delta$ and in the exterior of $T$,
	\item for some (arbitrary but fixed) point $o$ in $T$ each of the three rays emanating from $o$ through one of the corners of $T$ contains one of $a_i$, $b_i$, and $c_i$,
	\item $\chi(a_i,b_i,c_i)=\lt$,
	\item the straightline segments $x a_i$, $y b_i$, and $z c_i$ do not intersect each other.
\end{itemize}

\begin{figure}
	\centering
	\includegraphics{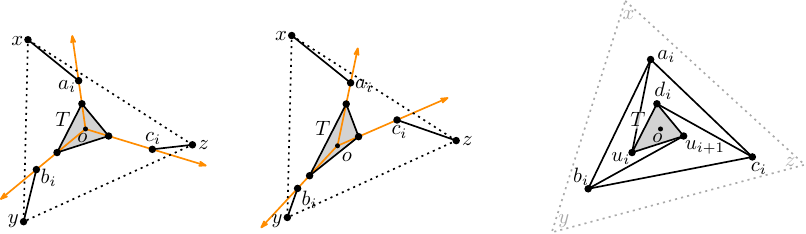}
	\caption{Left \& Middle: Placement of $a_i$, $b_i$, and $c_i$ on rays emanating from some point $o$ in $T$. Right: The edges $a_i u_i$, $b_i u_{i+1}$, and $c_id_i$ do not cross, no matter how $u_i$, $u_{i+1}$, and $d_i$ are arranged as the corners of $T$ (due the construction of $G_i$ depending on the value of $d_W(i)$).}
	\label{fig:SGE-innerTriangle}
\end{figure}

Such a placement is possible, since at least two sides of $\Delta$ are intersected by the chosen rays emanating from $T$, see \cref{fig:SGE-innerTriangle} (left and middle).
The placement on these rays ensures that for each of $a_i$, $b_i$, and $c_i$ the straightline segment to any vertex in $\{d_i, u_i, u_{i+1}\}$ does not intersect the interior of $T$.
Moreover, the neighbor of $a_i$ in $\{d_i, u_i, u_{i+1}\}$ comes, counterclockwise, before the neighbor of $b_i$ in $\{d_i, u_i, u_{i+1}\}$ which in turn comes before the neighbor of $c_i$ in $\{d_i, u_i, u_{i+1}\}$, which is due to the construction of $\widetilde{G}_i$ dependent on the value of $d_W(i)$.
This shows that the edges incident to $a_i$, $b_i$, or $c_i$ in $G_i$ neither intersect each other nor any other edges of $G_i$ drawn yet, see \cref{fig:SGE-innerTriangle} (right).
Finally, we place $p$ arbitrarily in the interior of $\Delta$.
If necessary, we apply a small perturbation, preserving the order types of non-collinear triples of points, to ensure that the vertices are mapped to distinct points and not mapped to the interior of edges.
With this placement, all graphs $\widetilde{G}_1,\ldots,\widetilde{G}_{t-2}$ and the non-primed subgraph of $G_\text{frame}$ are drawn without crossings.

\begin{figure}
	\centering
	\includegraphics{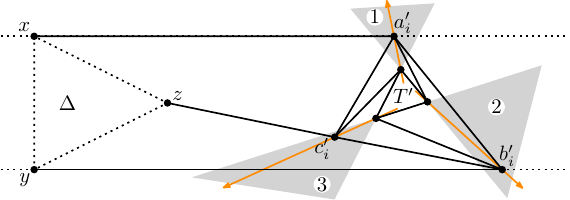}
	\caption{The placement of $a_i'$, $b_i'$, and $c_i'$ regarding $T'$. }
	\label{fig:SGE-outerTriangle}
\end{figure}

It remains to place the primed vertices, which we will place in the exterior of $\Delta$.
We start with the vertices from $S=\cup_{i=1}^{t-2} \{d'_i, e'_i, f'_i\} \cup \{u'_j\colon u_j\in V(W)\}$.
We embed $S$ such that its embedding forms a copy of the already constructed embedding of the corresponding non-primed vertices.
However, the embedding of $S$ is reflected along the y-axis, rotated so that no two points lie on a common horizontal line, and translated to the right such that each straight line through two points from $S$ intersects the horizontal lines through $x$ and $y$ at x-coordinates larger than $1$ (to the right of $\Delta$).
Note that $S$ is embedded strictly between the horizontal lines through $x$ and $y$.
Now we place, for each $i$, with $1\leq i\leq t-2$, the vertices $a'_i$, $b'_i$, and $c'_i$.
Consider the triangle $T'$ formed by $d'_i$, $u'_i$, and $u'_{i+1}$.
We call the corners of $T'$ the \emph{first}, \emph{second}, and \emph{third} corner such that the topmost corner is the first and the second corner follows clockwise along the boundary of $T'$.
The rotation above ensures that there is a unique topmost vertex.
We place $a'_i$ on the horizontal line through $x$ such that it lies on a ray which starts in the interior of $T'$ and runs through the first corner of $T'$ (that is, within the gray area labeled $1$ in \cref{fig:SGE-outerTriangle}).
We place $b'_i$ on the horizontal line through $y$ such that it lies on a ray which starts in the interior of $T'$ and runs through the second corner of $T'$ (that is, within the gray area labeled $2$ in \cref{fig:SGE-outerTriangle}).
We place $c'_i$ close to the third corner, and in the exterior of $T'$, such that it lies on a ray starting in the interior of $T'$ and running through the third corner (that is, within the gray area labeled $3$ in \cref{fig:SGE-outerTriangle}).
Finally, we place $p'$ at position $(2,0)$.
This placement ensures that
\begin{itemize}
	\item the triangle formed by $a_i'$, $b_i'$, and $c_i'$ lies to the right of $\Delta$  and has $T'$ in its interior (due to the placement within the gray areas and due to the translation of $S$ to the right of $\Delta$),
	\item $\chi(a_i',b_i',c_i')=\rt$ (due to placing the vertices clockwise around $T'$),
	\item the edges between $\Delta$ and  $\{a_i', b_i', c_i'\}$ do not intersect each other, intersect $\Delta$ only in their respective endpoint, and intersect the triangle formed by $a'_i$, $b'_i$, and $c'_i$ only in their respective endpoint (as $zc'_i$ lies horizontally strictly between the other two edges),
	\item the edges from $a_i'$, $b_i'$, and $c_i'$ to $T'$ neither intersect each other nor the interior of $T'$ (since $\chi(a_i',b_i',c_i')=\rt$ and $a_i'$, $b_i'$, and $c_i'$ are placed into the three gray regions distinctly).
\end{itemize}

If necessary, we apply a small perturbation to ensure that the vertices are mapped to distinct points and not mapped to the interior of edges while preserving the order types of non-collinear triples of points.
Altogether, we constructed a simultaneous crossing-free geometric drawing of $G_1,\ldots,G_{t+2}$, and of $G_\text{frame}$, as desired.

This reduction shows that \SGE\ is \ER-hard for edge-disjoint input graphs, since the Directional Walk Problem is \ER-hard for directional walks without repeated edges by \cref{thm:walk-1time}.
As \SGE\ is \ER-complete, the restricted variant is contained in \ER\ as well and hence \ER-complete.

\section{Conclusions}\label{sec:conclusions}

We showed that the simultaneous geometric embedding problem is \ER-complete for edge-disjoint input graphs.
Our proof is based on a reduction where the order type of a point set is encoded in a set of graphs such that the order type is realizable if and only the graphs admit a simultaneous geometric embedding.
Applying this reduction to a suitable partial order type yields a set of edge-disjoint graphs.

Realizability of partial order types seems an interesting topic on its own and we briefly discuss some observations next.
For a partial order type $\chi$ of some universe $U$ of points we consider the associated $3$-uniform hypergraph $\mathcal{H}_\chi$ with vertex set $U$ where the edges correspond to the triples of points with prescribed order type.
We have the following observations from an edge-density perspective:

\begin{itemize}
	\item If $\mathcal{H}_\chi$ is $2$-degenerate, that is, there is an ordering of $U$ such that each vertex is the last vertex in at most two edges of $\mathcal{H}_\chi$, then $\chi$ is realizable.
	
	Indeed, we embed the points one after the other in the mentioned order.
	We maintain the invariant, that the straight lines through the pairs of embedded vertices have pairwise distinct slopes (no two are parallel).
	For each point $u$ not embedded yet, there are at most two restrictions with respect to the already embedded points.
	Due to the embedding without parallel lines, there is always a placement of $u$ satisfying all restrictions and the invariant.
	Hence, we can embed all vertices.
	
	This case includes, for example, partial order types where each vertex appears in at most two triples ($\mathcal{H}_\chi$ has maximum degree $2$), directional walks without repeated vertices, and similarly defined directional trees and cycles.
	
	\item Realizability is \ER-hard to decide for partial order types $\chi$ with $\mathcal{H}_\chi$ of maximum degree at most $\binom{131}{2}$.
	
	This follows from a result of Schaefer~\cite[Theorem 1]{Sch21} as mentioned in the introduction.
	
	\item Realizability is \ER-hard to decide for partial order types $\chi$ with at most $\binom{n}{2}-1$ edges in~$\mathcal{H}_\chi$, where $n$ denotes the number of points in the universe.
	
	This follows directly from \cref{thm:walk-1time}, as directional walks without repeated edges have length at most $\binom{n}{2}+1$ (so at most $\binom{n}{2}-1$ inner vertices with prescribed order type).
\end{itemize}

Based on these observations we are interested in answers to the following questions.

\begin{enumerate}[{Question} 1:]
	\item Is there a family of partial order types such that the realizability problem for this family is \NP-complete?
	
	\item What is the complexity of the realizability problem for partial order types $\chi$ with $\mathcal{H}_\chi$ of maximum degree $3$?
	
	\item What is the complexity of the realizability problem for partial order types~$\chi$ where any two distinct edges in $\mathcal{H}_\chi$ intersect in at most one vertex (such hypergraphs are sometimes called simple)?
\end{enumerate}

Realizability seems also interesting for partially defined CC-systems or oriented matroids.
Deciding whether a partially defined oriented matroid (or chirotope) can be extended to an oriented matroid is known to be \NP-complete~\cite{Tschirschnitz01}.
Partially defined oriented matroids appear a few times in the literature, for instance in work by Bokowski and Richter~\cite{BR90} studying realizability.

\bibliographystyle{plainurl}
\bibliography{refs}

\end{document}